\documentclass[a4paper,reqno,nofootinbib,superscriptaddress,12pt]{iopart}

\usepackage{graphicx,color}
\usepackage{dcolumn}
\usepackage{epsfig}
\usepackage{ulem}

\usepackage{amsfonts}
\usepackage{iopams,setstack}
\usepackage{euscript}
\usepackage{amssymb}
\usepackage{amsthm}
\usepackage{newlfont}
\usepackage{mathrsfs}
\usepackage{subfigure}
\usepackage{srcltx}

\newcommand{\opunit}{\text{1}\kern-0.22em\text{l}}

\newcommand{\id}{\textrm{d}}

\def\bea{\begin{eqnarray}}
\def\eea{\end{eqnarray}}
\def\ba{\begin{array}}
\def\ea{\end{array}}
\def\n{\nonumber}

\def\la{\langle}
\def\ra{\rangle}

\begin{document}

\title{Mobility transition in a dynamic environment}
\author{Urna Basu, Christian Maes }
\address{Instituut voor Theoretische Fysica, KU Leuven, Belgium}
\ead{urna.basu@fys.kuleuven.be}

\begin{abstract}
Depending on how the dynamical activity of a particle in a random environment is influenced by an external field $E$,  its differential mobility at intermediate $E$ can turn negative. We discuss the case where for  slowly changing random environment the driven particle shows negative differential mobility while that mobility turns positive for faster environment changes. We illustrate this transition using a 2D-lattice Lorentz model where a particle moves in a background of simple exclusion walkers. The effective escape rate of the particle (or minus its collision frequency) which is essential for its mobility-behavior depends both on $E$ and on the kinetic rate $\gamma$ of the exclusion walkers. Large $\gamma$, i.e., fast obstacle motion, amounts to merely rescaling the particle's free motion with the obstacle density, while slow obstacle dynamics results in particle motion that is more singularly related to its free motion and preserves the negative differential mobility already seen at $\gamma=0$.  In 
more general terms that we also illustrate using one-dimensional random walkers, the mobility transition is between the time-scales of the quasi-stationary regime and that of the fluid 
limit.
\end{abstract}


\maketitle

\section{Introduction}
The differential mobility $\mu(E)$ of a colloid in a medium measures how for a given external field $E$ its velocity increases with changing $E\rightarrow E+\id E$.   A well--known implication of the fluctuation--dissipation theorem for an equilibrium environment at temperature $T$ is the Sutherland-Einstein relation $D(E=0) = T\,\mu(E=0)$ between the diffusion constant $D$ at zero field and the mobility $\mu$,  \cite{Einstein}.  Going to higher $E$-values, this equality fails \cite{mod_Einstein} and even more interestingly, it easily happens that $\mu(E) \leq 0 $ implying that pushing harder slows down the particle \cite{zia,sog}.  A possible unifying framework for understanding these negative differential mobilities (and other conductivities \cite{Li}) has been found in the study of nonequilibrium response \cite{neg_resp}.  There enters the dynamical activity which is a measure of time--symmetric currents and expresses a ``nervosity'' of the particle during its motion. The point is that the nonequilibrium 
contribution to the response modifies the Sutherland--Einstein (or more generally, the Kubo) formula by subtracting from $D(E)$  the time--correlation function between the velocity of the particle and its differential dynamical activity, which can be large.  We have called it the \emph{frenetic} contribution \cite{fren} to contrast it with the \emph{entropic} contribution which makes the standard Green--Kubo or Helfand term, \cite{Green1954,Kubo1957,helf}.

The present paper  adds an extra parameter $\gamma$ in the study of the differential mobility $\mu(E,\gamma)$, denoting in general by $\gamma$ an inverse time-scale or the rate at which constituents of the environment are themselves moving. We consider here single particles moving in a bath of obstacles. This is an example of the more general study of random walks in a dynamic random environment; see e.g. \cite{RWSEP, RWRE} in the mathematics literature. Parameters like $\gamma$ can refer to the changing architecture or geometry of the environment, or they can effectively parametrize the dynamics of the particle. In all cases here we assume that the dynamics of the environment is itself undriven for fixed particle position. Specific examples follow in the next sections.  For broader physics background on the models and for previous work we also refer to \cite{os1,os2,os3,Franosch}.\\

 The scenario is always that $\mu(E,\gamma=0) \leq 0$ for some range of $E\geq E_c(0)>0$, where $\gamma=0$ indicates a static (or quenched) random environment. The case $\gamma=0$ is however a singular limit.  Our main study and observation is that the negativity 
 $\mu(E,\gamma) \leq 0, E\geq E_c(\gamma)>0$ is stable for  $0\leq \gamma \leq \gamma_c$ after which, for $\gamma >\gamma_c$, the differential mobility is positive $\mu(E,\gamma) > 0$ for all fields $E$.  In other words, there is a finite threshold field $E_c(\gamma)$ (only) for $\gamma< \gamma_c$.   The critical value $\gamma_c >0$ first of all marks the regime of \underline{quenched} static random environment and the regime of annealed random environment.  Indeed, taking the time-scale at which the driven particle moves to be of order one for all values of the field $E$, we have that $\gamma \gg 1$ corresponds to large time-separation under which the particle moves in an averaged--out environment where $\mu(E,\gamma \gg 1)
>0$.  That can be called the \underline{fluid limit}.  Secondly, the regime $\gamma < \gamma_c$ is \underline{quasi-stationary} for the particle, where the relaxation time $\tau(E,\gamma)$ for the particle dynamics is still much smaller than the time-scale $1/\gamma$ for the environment.  In other words, negative response is expected to hold when $\tau(E,\gamma) \ll \gamma^{-1}$ for $E\geq E_c(0)$. If $\tau(E,\gamma)\leq \tau(E_c(0),0)$, it suffices that $\tau(E_c(0),0) \ll \gamma^{-1}$ which implies that $\gamma_c\ll 1$ is very small when $\tau(E,\gamma=0)$ grows fast with the field $E$.\\

The plan of the paper is as follows.  We start in the next section with the general question in the context of integrating out a dynamic environment.  Section \ref{sec:log} is devoted to  a computational and numerical analysis of the lattice Lorentz model with moving obstacles as random dynamic environment.  We describe the regime of negative differential mobility $\mu(E,\gamma)$ where $E$ is the driving field on the particle and $\gamma$ is the rate at which the individual obstacles move.  Section \ref{sec:one} is devoted to a number of toy-models of walkers in quasi-one dimensional architectures  that enable easier access for  understanding the mobility transition. 

\section{General question}\label{sec:gq}
A tagged degree of freedom (called, particle position) is subject to a constant external driving $E$  and moves in a dynamical environment (called, obstacles).  We do not assume that the coupling is ``weak,'' or that the particle is ``small'' or that its motion is ``slow'' with respect to the obstacles.  Particle and obstacles are coupled and are in weak contact with an equilibrium bath at a fixed inverse temperature $\beta$ that we set to one.  The obstacles are undriven and the whole process is reversible at $E=0$.  For fixed particle position the obstacles can be considered to be an equilibrium fluid specified by some density $\rho$ and inverse temperature $\beta=1$.  Calling  $\gamma$ a measure for the time-scale separation between obstacle and particle dynamics, we have $\gamma\uparrow\infty$ corresponding to that equilibrium fluid limit.   On the other hand, for $\gamma\downarrow 0$ the particle sees a quasi--static random environment.\\

  We consider the regime where the motion of the undriven particle ($E=0$ but arbitrary $\gamma$) is asymptotically diffusive with a linear response regime in $E$ satisfying the Kubo or Sutherland-Einstein relation.  The \underline{general question} is to understand the dependence of the particle velocity {\it versus} field $E$ also for intermediate and large $E$ as a function of $\gamma$.  Selecting the direction of that field as the one--dimensional space for the particle's position, one version of the question boils down to understanding the effective particle's motion under coarse--graining, i.e., after integrating out the environment.  To characterize that effectively one--dimensional motion and for simplicity assuming a Markovian random walk we essentially need to know the effective rate of moving forward ($p_{{\text{eff}}}$) and that of moving backward ($q_{{\text{eff}}}$).
 A  more physical parametrization is contained in two types of information:\\
(1)  Because of the undriven character of the environment, we expect that the effective biasing field on the particle is essentially still given by $E$.  From the hypothesis of local detailed balance \cite{kls} we would then get
\bea
\beta_{{\text{eff}}}\,E = \log\frac{p_{{\text{eff}}}}{q_{{\text{eff}}}} \label{eq:b_eff}
\eea
where it cannot be excluded entirely that the effective temperature   $\beta_{{\text{eff}}} \neq \beta (=1)$ because of possible entropy fluxes between particle and obstacles at some different effective temperature depending on $\gamma$ and $E$.  If however the obstacle density $\rho$ is low or $\gamma$ is large, we expect $\beta_{{\text{eff}}} = \beta=1$ to very good approximation.\\
(2) The effective escape rate 
\begin{equation}\label{eq:star}
g(E,\gamma) = p_{{\text{eff}}}+ q_{{\text{eff}}}
\end{equation}
measures the dynamical activity of the particle in terms of the time--symmetric current along the direction of the field.\\

Taking that picture of the one--dimensional walker seriously and combining (\ref{eq:b_eff}) with (\ref{eq:star}) we would obtain a speed of the particle in the direction of the field 
specified as
\begin{equation}\label{eq:vr}
\langle v\rangle = p_{{\text{eff}}} - q_{{\text{eff}}} = \frac{1- e^{-\beta_{{\text{eff}}}\,E}}{{1+ e^{-\beta_{{\text{eff}}}\,E}}}\,g(E,\gamma)
\end{equation}
The primary question addressed here is whether it is possible to have a non-monotonic dependence of $\la v \ra$ on the field $E$ or in other other words, whether this system shows negative differential mobility and for what $\gamma.$ For the linear behavior around $E=0$ equation (\ref{eq:vr}) gives $\langle v\rangle = \beta_{{\text{eff}}}\,E\,g(0,\gamma)$ or $\mu(E=0,\gamma) = \beta_{{\text{eff}}}\,g(0,\gamma)>0$. For intermediate and large $E$ in (\ref{eq:vr}) the variation of the speed with $E$ crucially depends on how $g(E,\gamma)$ behaves with $E$ for a given $\gamma$.  In particular, taking now $\beta_{{\text{eff}}}=1$, if $g(E,\gamma)$ decreases fast enough with the driving field $E$, then it is possible to have such a scenario. As $g(E,\gamma)$ ultimately refers to the activity of the particle, we are led to the question  whether the particle gets more or gets less `trapped' by increasing $E$. In the next section we focus on a specific model system, namely lattice Lorentz gas with moving obstacles, to 
investigate these questions.

\section{Model}\label{sec:log}

\begin{figure}[h]
 \centering
 \includegraphics[width=8 cm]{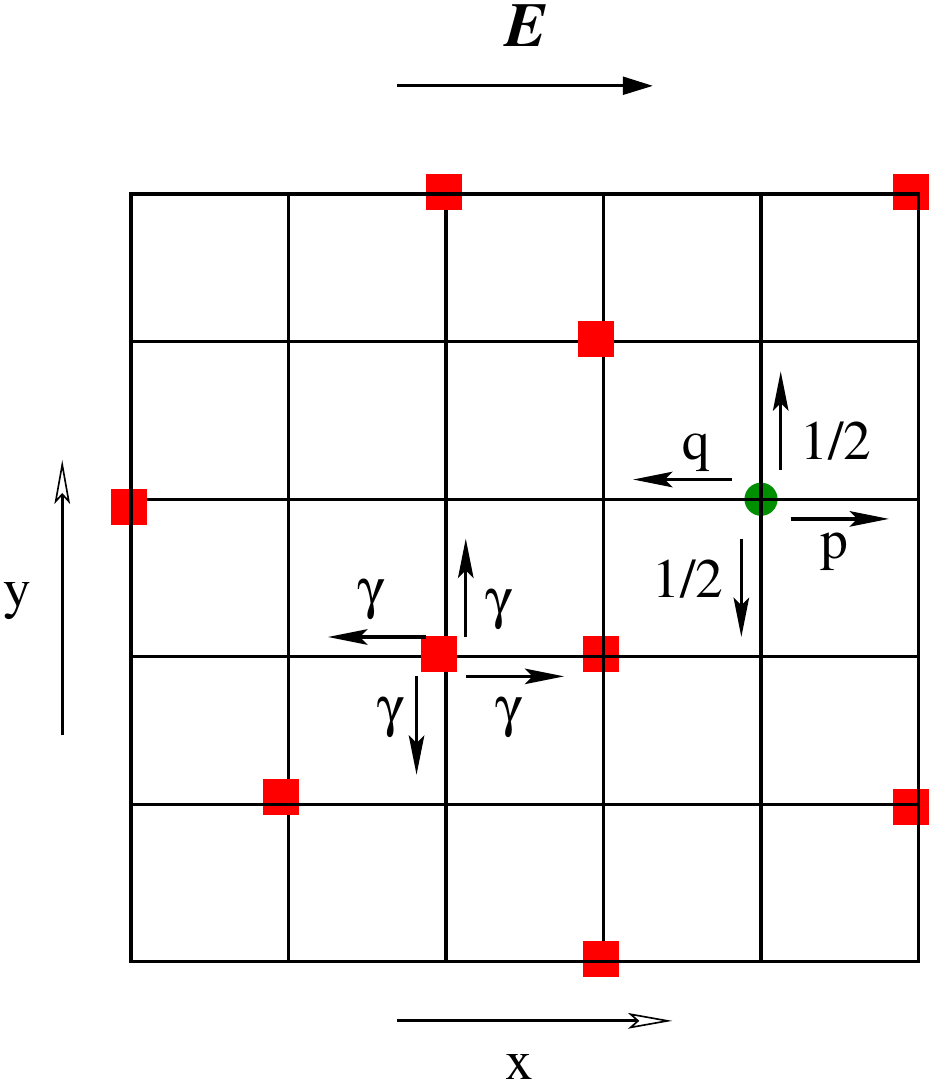}
 \caption{Single driven charged particle (green circle) interacting via exclusion with symmetrically hopping sea particles (red squares).}
 \label{fig:model}
\end{figure}

Let us consider ${\cal N}$ particles distributed on a 2D square lattice of size $L \times L.$   We assume periodic boundary conditions and the density $\rho = {\cal N}/L^2$ does not evolve in time.   These particles, referred to as the `obstacles' or `sea particles,' perform a symmetric simple exclusion process with rate $\gamma$. There is another `charged' or `probe' particle in the system which is driven by an `electric' field $E$ pointing in the $+x$ direction. Consequently  this charged particle performs a biased random walk in the $x$-direction --- it moves to the right (left) with rate $p(q)$  while the motion along the $y$-direction is symmetric --- the rates of moving up and moving down are equal to $1/2$ (see Fig.~\ref{fig:model}). We assume local detailed balance \cite{kls} so that $p/q=e^{E}.$ The charged particle interacts with the sea particles via exclusion --- it can move only when the target site is empty; the sea particles act as `obstacles' for the charged one. This 
model can thus
be considered as a lattice Lorentz gas with moving obstacles, \cite{Lorentz}. 

A discrete time version of the model is studied in \cite{os1}; here we work in continuous time where the jumps have rates which follow the condition of local detailed balance \cite{kls}.
The local detailed balance condition does not specify the rates completely however; we choose 
\bea\label{eq:cho}
p = {e^{E/2}\over e^{ E/2}+e^{- E/2}},\quad\;\; q = {e^{- E/2}\over e^{ E/2}+e^{- E/2}}
\eea
so that in the absence of obstacles the escape rate in the horizontal direction is $p+q=1.$   In that way, we do not impose by hand a specific change in escape rate for varying field $E$ at zero obstacle density  (free motion).  Note that $p \simeq 1- e^{-E}, q \simeq e^{-E}$ for large $E$.\\

 To numerically simulate this system, we use a {\it random sequential} updating scheme. First a site is selected randomly - if it is occupied by the probe particle or a sea particle then it attempts to jump to one of the four neighbouring sites with corresponding rates, the attempt being successful only when the target site is empty. On the other hand if the selected site is vacant then nothing is done. In this scheme each update, whether successful or not,  corresponds to an increase in time $\id t=1/L^2.$ \\

Let $v$ denote the net number of steps per unit time that the charged particle makes in the $+x$-direction.  An expression for the average speed $\langle v\rangle$ can be obtained from the dynamical updating rules. Let us denote the instantaneous position of the charged particle by $(X_t,Y_t).$ Also let $\eta(x,y)$ denote the occupation number for the sea particles,  $i.e.$, $\eta(x,y)=1,0$ depending on whether site $(x,y)$ is occupied by a sea particle or not. Then, the dynamics gives
\bea
\la v \ra =\frac d{dt} \la X_t \ra &=& p \la (1- \eta_t(X_t+1,Y_t)) \ra - q \la (1- \eta_t(X_t-1,Y_t)) \ra 
\eea
The expectation $\la \eta_t(X_t+1,Y_t) \ra$ measures the probability (or fraction of time) that the right neighbouring site of the charged particle is occupied, and similarly for $\la \eta_t(X_t-1,Y_t) \ra$.  Or, with $t_R(t_L)$ denoting the total time in $[0,t]$ during which there is an obstacle  to the right (left) neighbouring site of the charged particle, we have under stationary conditions that
\bea
\la v \ra &=& p(1- \la t_R \ra /t ) - q(1- \la t_L \ra /t ) \cr
&=& p_{{\text{eff}}} - q_{{\text{eff}}} \label{eq:vpqeff}
\eea
where we identify, $p_{{\text{eff}}}=p(1- \la t_R \ra /t), q_{{\text{eff}}}= q(1- \la t_L \ra /t ).$ 
That is the speed of a biased one--dimensional random walker with effective rates $p_{{\text{eff}}},q_{{\text{eff}}}$ and we are exactly realizing the spirit of formula (\ref{eq:vr}).  Following these ideas  we then expect negative differential mobility when the effective escape rate $ p_{{\text{eff}}} + q_{{\text{eff}}}$ decreases sufficiently fast as a function of $E$.\\

\begin{figure}[t]
 \centering
 \includegraphics[width=7 cm]{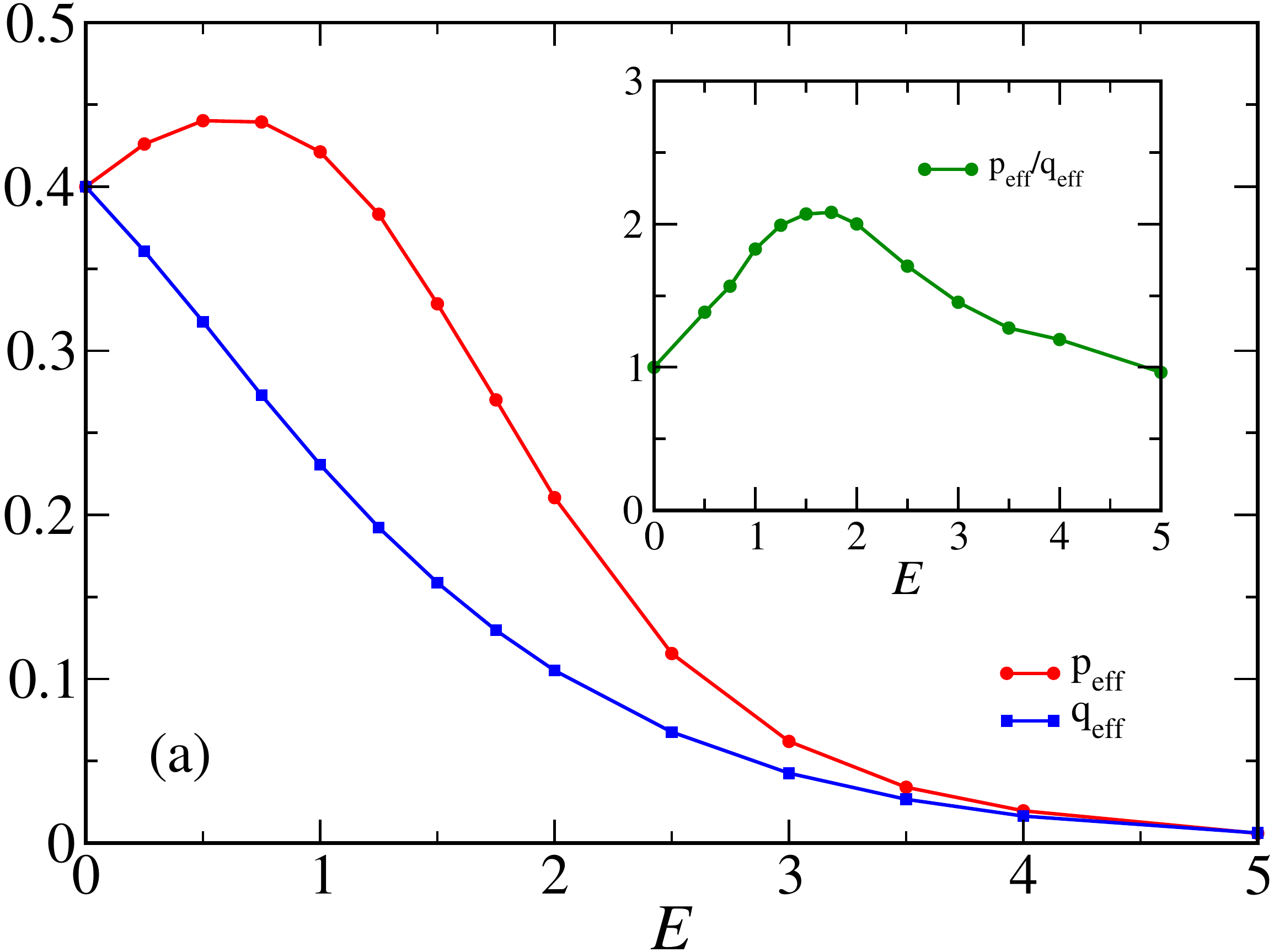}\hspace*{0.6 cm}\includegraphics[width=7 cm]{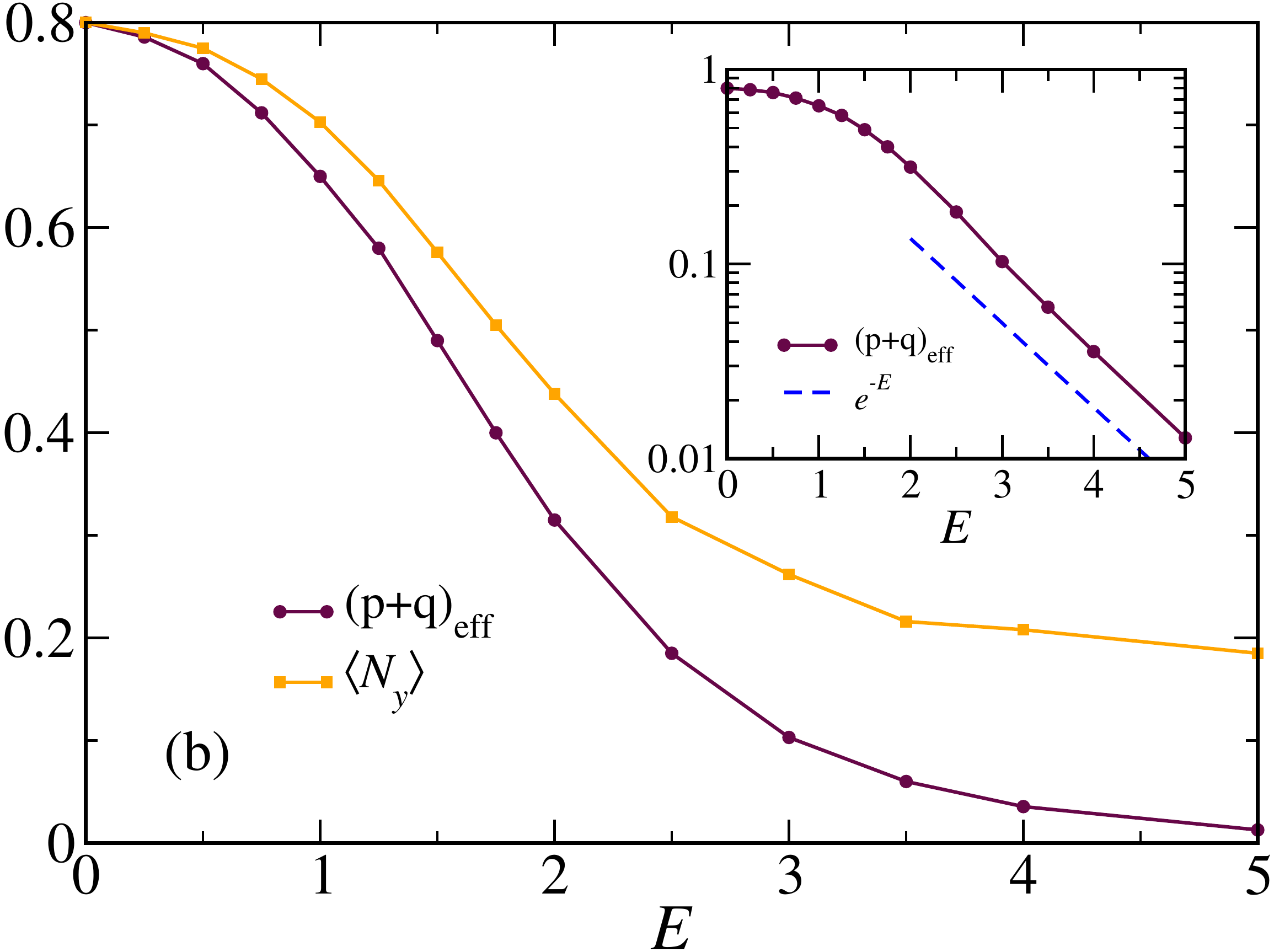}
 \caption{Case $\gamma=0.$ (a) The effective jump rates $p_{\text{eff}}$ (red circles) and $q_{\text{eff}}$ (blue squares) as a function of field $E.$ Inset shows the effective bias $p_{\text{eff}}/q_{\text{eff}}$ for the same. (b) The effective escape rate in $x$ and $y$ directions $\la N \ra/t=(p+q)_{\text{eff}}$ (dark maroon circles) and $\la N_y \ra$ (light orange squares) respectively. The inset compares the former with $e^{-E},$ shown in dashed line,  in semi-log scale. The system used is of size $100 \times 100$ and density of the sea particles $\rho=0.2.$ } 
 \label{fig:g0_pq}
\end{figure}

\subsection*{Static Obstacles: $\gamma=0$}

The $\gamma=0$ case corresponds to the driven lattice Lorentz gas and is known to show negative response irrespective of the choice of $p,q$ \cite{neg_resp,Franosch,Dhar}.   The study of this system was started in the context of diffusion in random media  with focus on the phase transition to a zero--current regime \cite{Dhar,Barma,Stauffer,Pandey}. Here we concentrate on the origin of the non--monotonicity of the velocity as a function of the field to understand the stability of the phenomenon for $\gamma>0$.\\

\begin{figure}[t]
 \centering
 \includegraphics[width=7.3 cm]{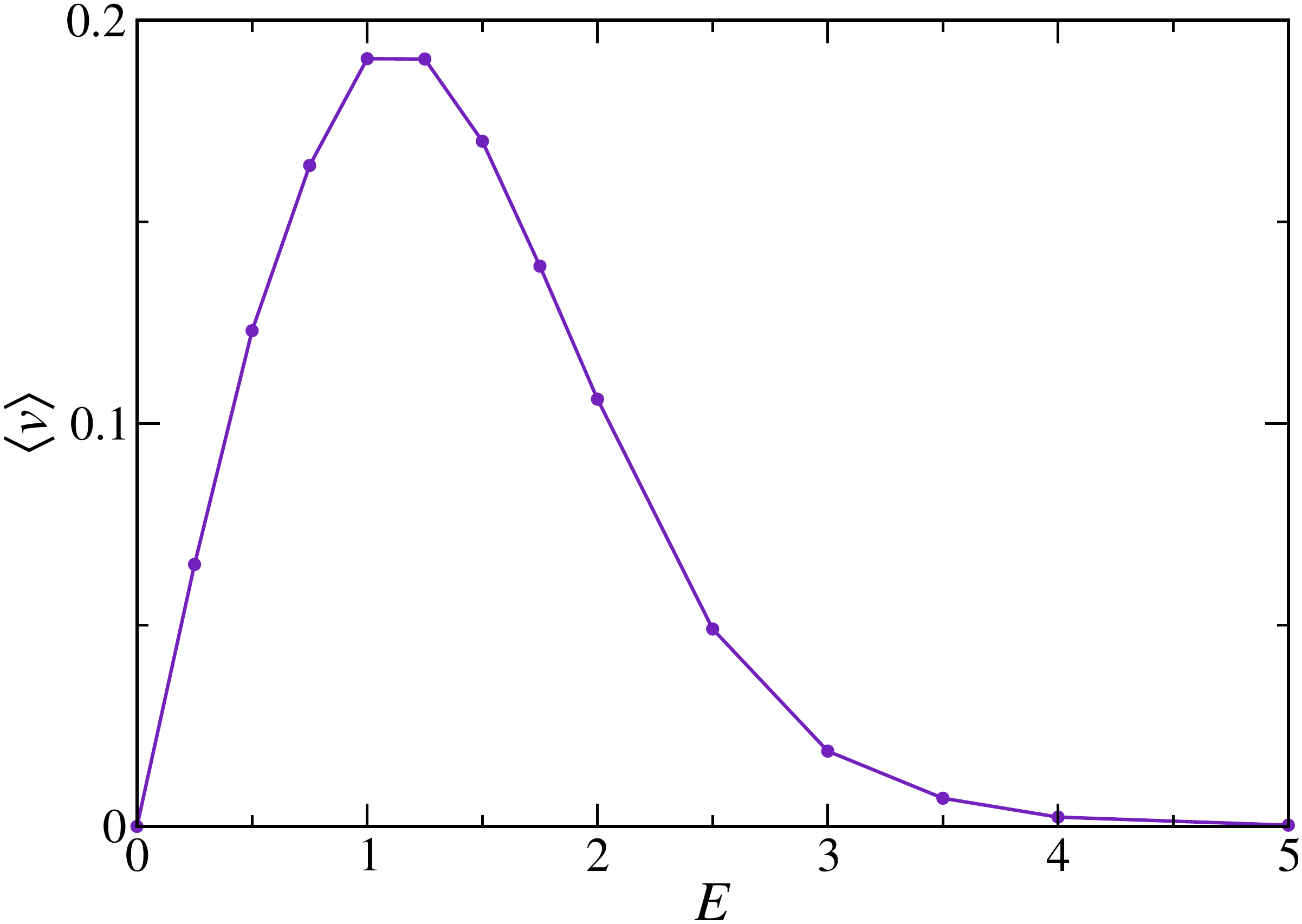}\hspace*{0.1 cm}\includegraphics[width=7.0 cm]{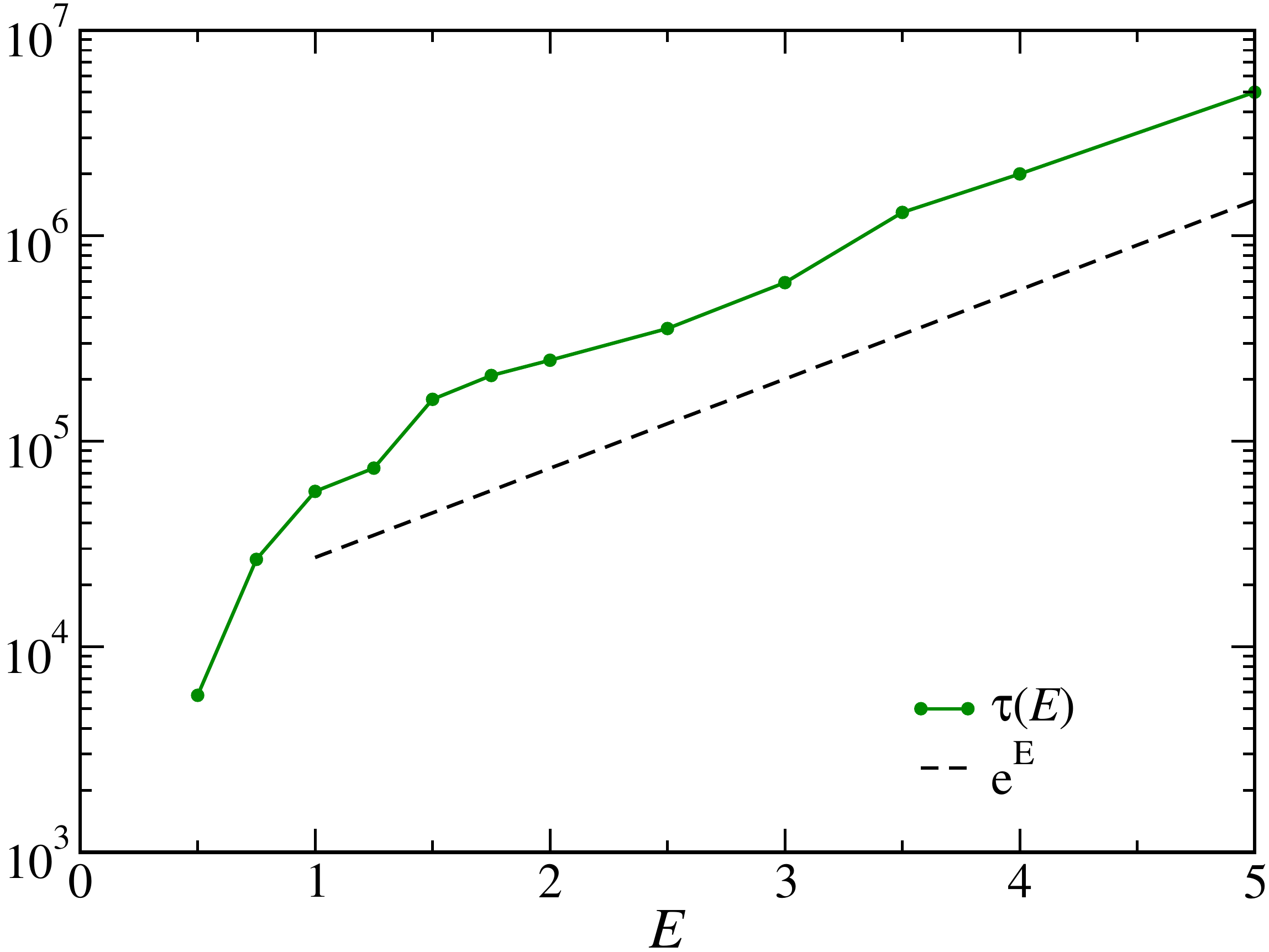}
 \caption{Case $\gamma=0.$ (a) $\la v\ra$ versus $E;$ negative differential response beyond $E>E_c \sim 1.25.$  (b) The relaxation time $\tau(E)$ increases exponentially with field $E.$ The density $\rho=0.2.$ }
 \label{fig:g0}
\end{figure}

The effective rates $p_{\text{eff}}$ and $q_{\text{eff}},$ as defined in (\ref{eq:vpqeff}), are plotted as a function of the field $E$ in Fig. \ref{fig:g0_pq}(a). The effective bias measured from the ratio $p_{\text{eff}}/q_{\text{eff}}$ is shown in the inset. The  non--monotonic nature of this curve indicates a strong violation of the local detailed balance in the steady state. Fig.~\ref{fig:g0_pq}(b) shows a plot of the sum of these quantities $(p+q)_{\text{eff}}$ (dark circles) which is a measure  of the dynamical activity. The inset suggests that this effective escape rate is exponentially decaying for large field $E.$ 
 As the stationary velocity is given by (\ref{eq:vr}), and both the factors $\frac{1- q_{\text{eff}}/p_{\text{eff}}}{1 + q_{\text{eff}}/p_{\text{eff}}}$ and $(p+q)_{\text{eff}}$ are decreasing for large $E$, the differential mobility becomes strongly negative.  Fig. \ref{fig:g0} (a) shows  the stationary speed $\la v\ra$ as a function of the field $E;$ it shows a strong negative differential mobility for $E>1.25=E_c(\gamma=0).$ \\
Interestingly, the activity in the $y$-direction as quantified by the average number of vertical jumps per unit time $\la N_y \ra$ is also decreasing with the field; see the upper curve (light orange squares) in Fig. \ref{fig:g0_pq}(b).  In contrast with the free motion in 2D the motion in the $x$- and in the $y$-directions are no longer independent.  
 
  It should be emphasized however that the $\gamma=0$ situation is a somewhat singular limit.  Various correlation functions do not reach a stationary value (at least within feasible computational time) for large fields. The relaxation time $\tau(E)$,  defined as the time required  to reach a stationary velocity,   increases exponentially with the field $E;$ Fig. \ref{fig:g0}(b) appears to verify $\tau(E) \sim e^E$ at least for large $E.$   
  Secondly and related,  the stationary distribution of the charged particle depends on the obstacle configurations for $\gamma=0$. All the simulations that we present here are thus averaged over many (from about 600 for large fields to 1000 for small fields) obstacle configurations, with 100 trajectories for each configurations. The lattice considered is a square of size $100 \times 100$ with periodic boundary conditions in both $x,y$ directions.

\subsection*{Moving obstacles : $\gamma \ne 0$ }

\begin{figure}[t]
 \centering
 \includegraphics[width=7.5 cm]{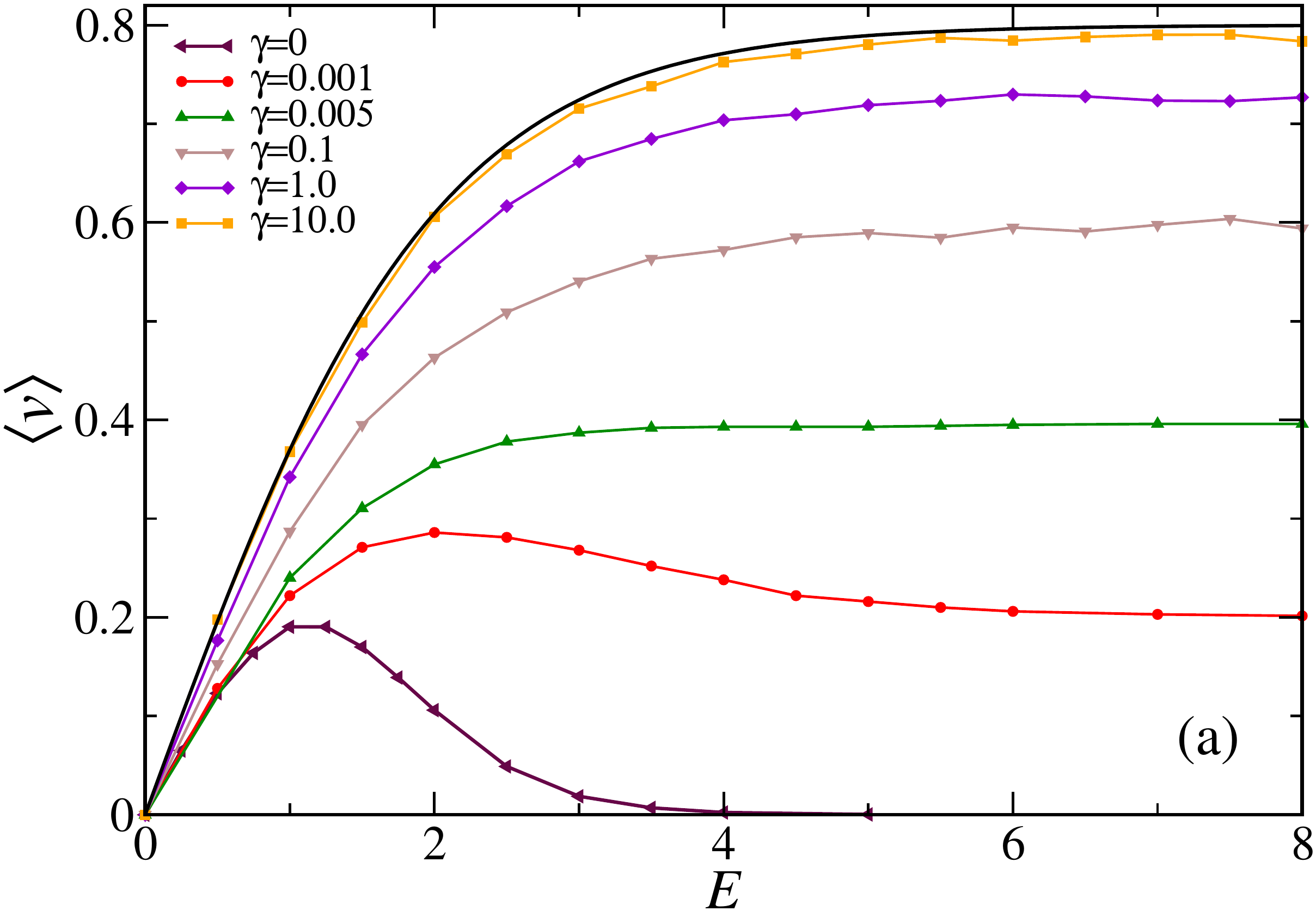} \hspace*{0.2 cm} \includegraphics[width=7.5 cm]{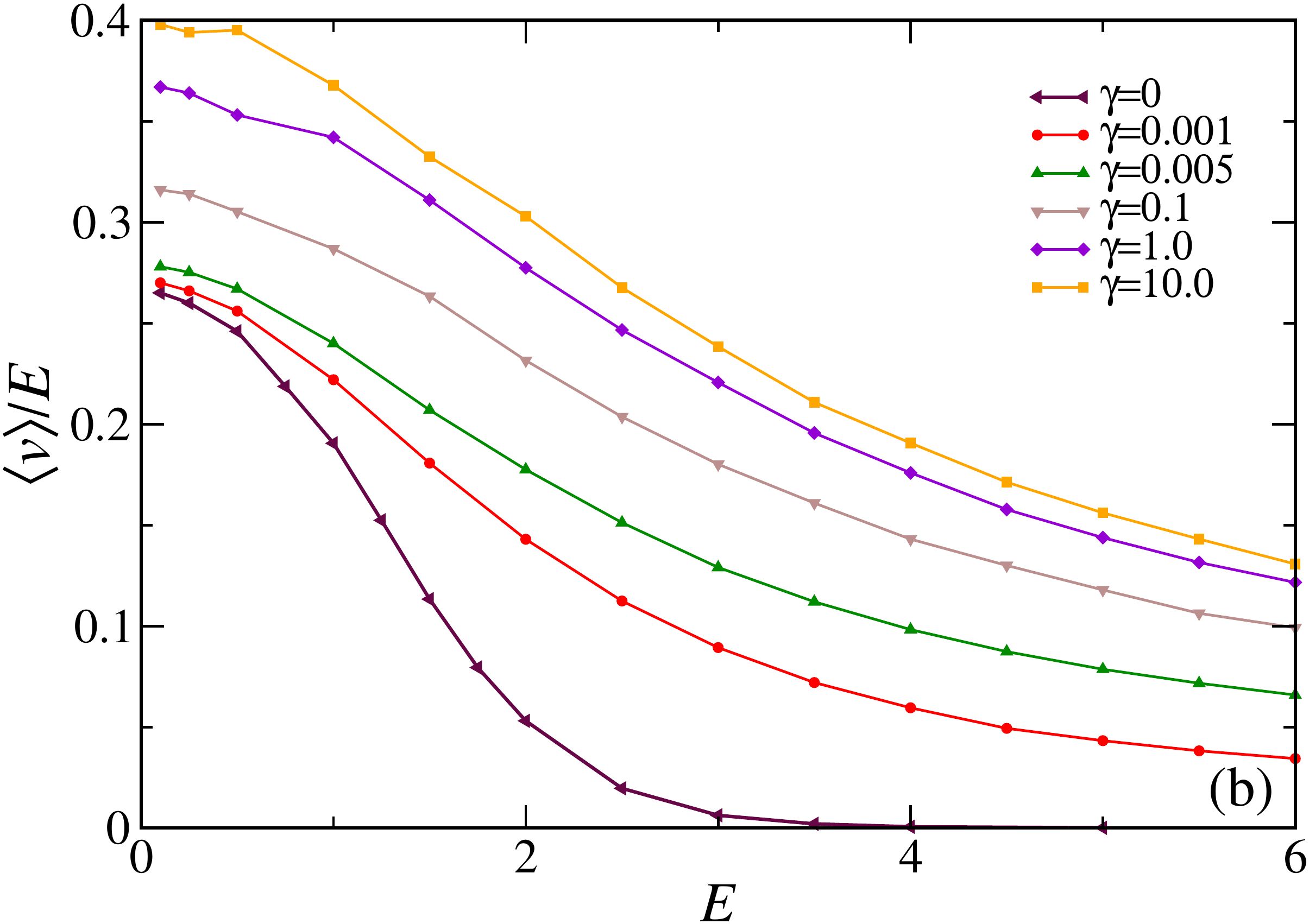}
 \caption{(a) $\la v \ra$ as a function of field $E$ for different values of $\gamma$  ranging from $0.001-10.0$ and density $\rho=0.2.$ For large $\gamma$ the current asymptotically reaches $(p-q)(1-\rho),$ shown in solid black line.  (b) Absolute mobility $\la v \ra/E$ as a function of the field $E$ for the  same parameters as in (a). The negative differential response can only be guessed from the steeper descent for the $\gamma \le 0.001$ curves. 
 }\label{fig:r0.2}
\end{figure}

\begin{figure}[h]
 \centering
 \includegraphics[width=7.5 cm]{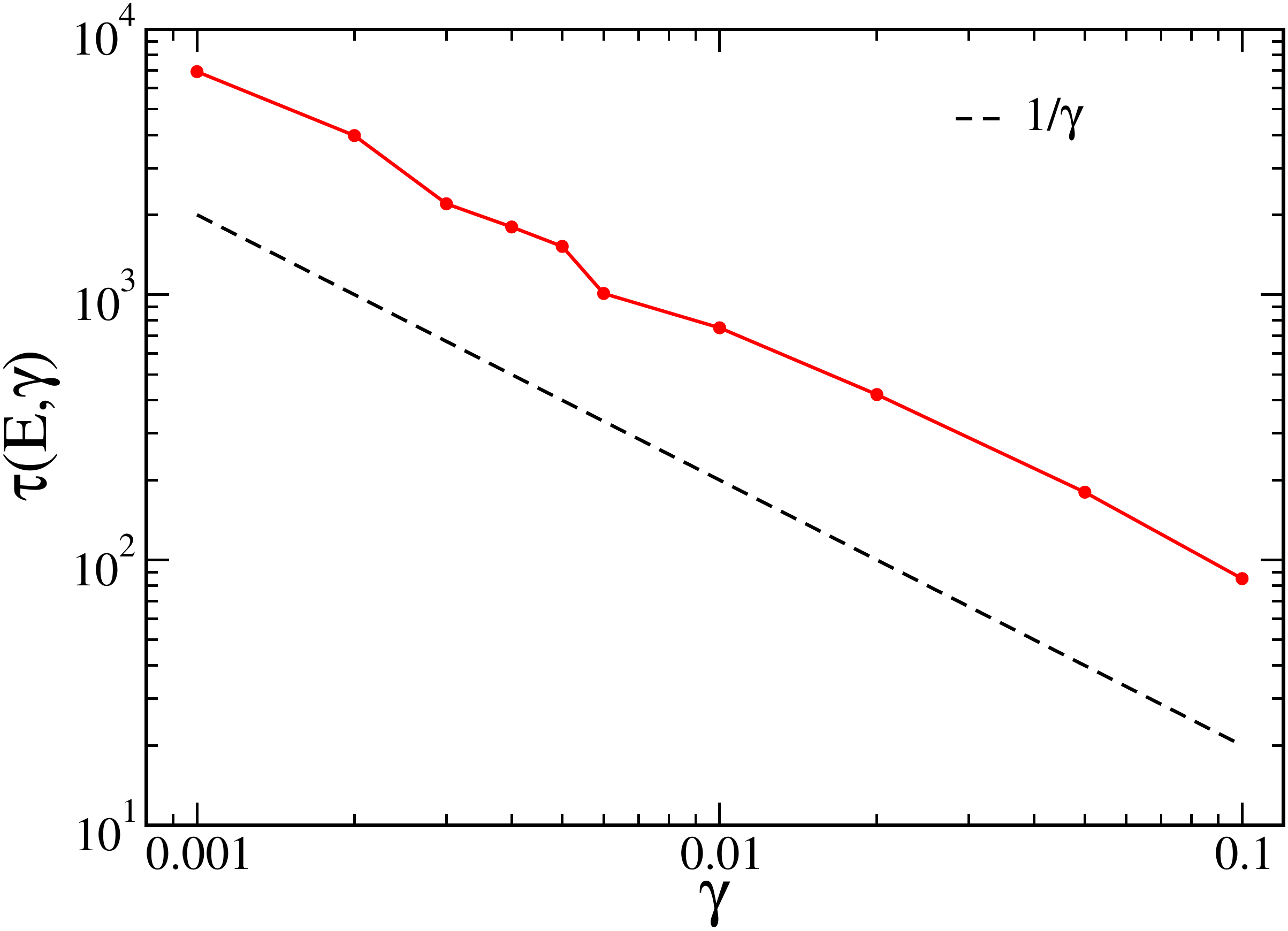}
 \caption{ Relaxation time $\tau(E,\gamma)$ as a function of $\gamma$ for a fixed $E=3.0.$ The dashed line corresponds to $\gamma^{-1}.$
 }\label{fig:r0.2_trlx}
\end{figure}

The non--zero $\gamma$ presents a dynamic environment for the charged particle which leads to a unique stationary behaviour and the charged particle's position is uniformly distributed over the lattice. Nevertheless, one can expect a negative differential response  
for `reasonably' small $\gamma$, {\it i.e.}, when the time scale associated to the charged particle is small compared to the sea particles. This picture is verified in Fig. \ref{fig:r0.2}(a) where we have plotted the dependence of the average speed $\la v \ra$ on the external field $E$ for different values of $\gamma$ ranging from $\gamma=10^{-3}$ to $\gamma=10$ for $\rho=0.2.$ The negative response is seen for very small values of $\gamma$ like $\gamma \sim 10^{-3}.$ The average speed becomes monotone increasing in $E$ for $\gamma>\gamma_c\simeq 0.005$ (dark green triangles in Fig. \ref{fig:r0.2}(a)).  As $\gamma$ is increased further the average speed approaches $\la v\ra^{\gamma \uparrow \infty} = (1-\rho)(p-q)$ (solid black line) as is expected in the fluid limit. Instead of looking at the differential mobility, one can also consider the  absolute mobility, defined as the ratio of average velocity and the applied field $\la v\ra/E,$ shown in Fig. \ref{fig:r0.2}(b). The initial constant region 
corresponds to the linear response (Kubo) regime.  Clearly, the presence of the transition in differential mobility is not directly reflected from the behaviour of ordinary mobility. The results remain qualitatively similar for other values of the density $\rho$; the absolute value of the current decreases of course as one goes to higher density. One difference with $\gamma=0$ is the limit $E\uparrow \infty$ for the average speed, saturating 
at a non-zero $\gamma$-dependent value whenever $\gamma \neq 0$ (see (\ref{eq:v_toy}) below). \\
The relaxation time of the charged particle now depends on both $E$ and $\gamma$ and we expect $\tau(E,\gamma) \sim \gamma^{-1}$ for a fixed $E$ and large $\gamma$ as can be verified from Fig. \ref{fig:r0.2_trlx}.  On the other hand, the  relaxation time for \emph{fixed} $\gamma>0$ increases exponentially with the field $E$ (not shown). \\

Similar to the $\gamma=0$ case we measure the effective dynamical activity and effective bias from the stationary values of $p_{\text{eff}}$ and $q_{\text{eff}}.$ Fig. \ref{fig:pqeff}(a) shows the dependence on $E$ of that effective escape rate $(p+q)_{\text{eff}}$
for different small values of $\gamma;$ for the sake of comparison we have added the escape rates for the $\gamma=0$ case too. Clearly the effective escape rates behave similarly in all these cases, decreasing in $E$. However, only for small $\gamma \sim 0.001$ the decay is fast enough to result in a non-monotonic velocity implying negative differential mobility.  The corresponding effective biasing field behaves as  $p_{\text{eff}}/q_{\text{eff}} \sim e^{E}$ as shown in Fig.~\ref{fig:pqeff}(b) thus maintaining the local detailed balance. In fact the effective temperature $\beta_{\text{eff}},$ defined in (\ref{eq:b_eff}) remains equal to $\beta=1$ at least for large field $E.$

\subsection{Response : Differential Mobility}

The differential mobility is the linear response coefficient of the speed to the perturbation $E \rightarrow E + \id E.$
The corresponding response formula is most conveniently derived using the path-integration formalism of \ref{ap}.  We give here the result for the choice (\ref{eq:cho}).\\
For a general path-observable $O$ of the system over a time-interval $[0,t]$  we find
\begin{equation}\label{eq:ree}
\frac \id {\id E} \la O \ra  =  \frac 12 \la J;O \ra -  p q \la (t_L-t_R);O \ra -\frac12 (p-q) \la N ; O \ra  
\end{equation}
where $\langle \cdot \,;\,\cdot\rangle$ denote connected correlation functions. Here, $N=N_\rightarrow + N_\leftarrow$ denotes the total number of jumps (right or left) made by the charged particle in time $[0,t]$ along the $x$-direction; $J=N_\rightarrow -N_\leftarrow$ is the total current and the speed is $v=J/t.$  The observables $t_L$ and $t_R$ are the dwelling times as defined through (\ref{eq:vpqeff}) in the previous section. Choosing $O=1$ in (\ref{eq:ree}) gives
\begin{eqnarray}\label{eq:Jpq}
\la J \ra &=& 2 pq \,\la t_L-t_R \ra + (p-q)\,\la N \ra \\
&=& \frac{2}{(e^{E/2} + e^{-E/2})^2}\,\la t_L-t_R \ra + \la N\ra\,\tanh\frac{E}{2} \nonumber
\end{eqnarray}

\begin{figure}[t]
 \centering
 \includegraphics[width=7 cm]{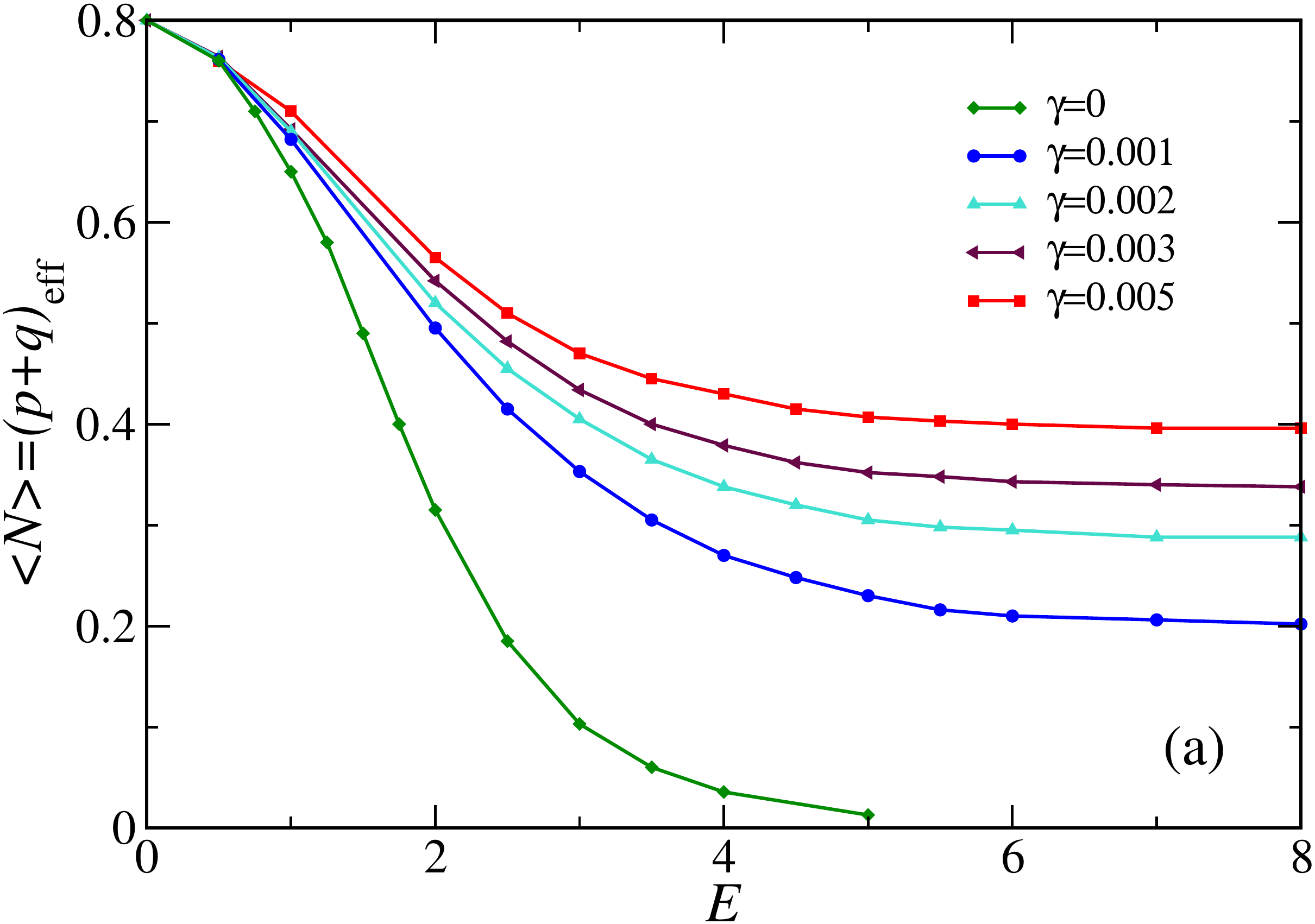}  \hspace*{0.2 cm}  \includegraphics[width=7 cm]{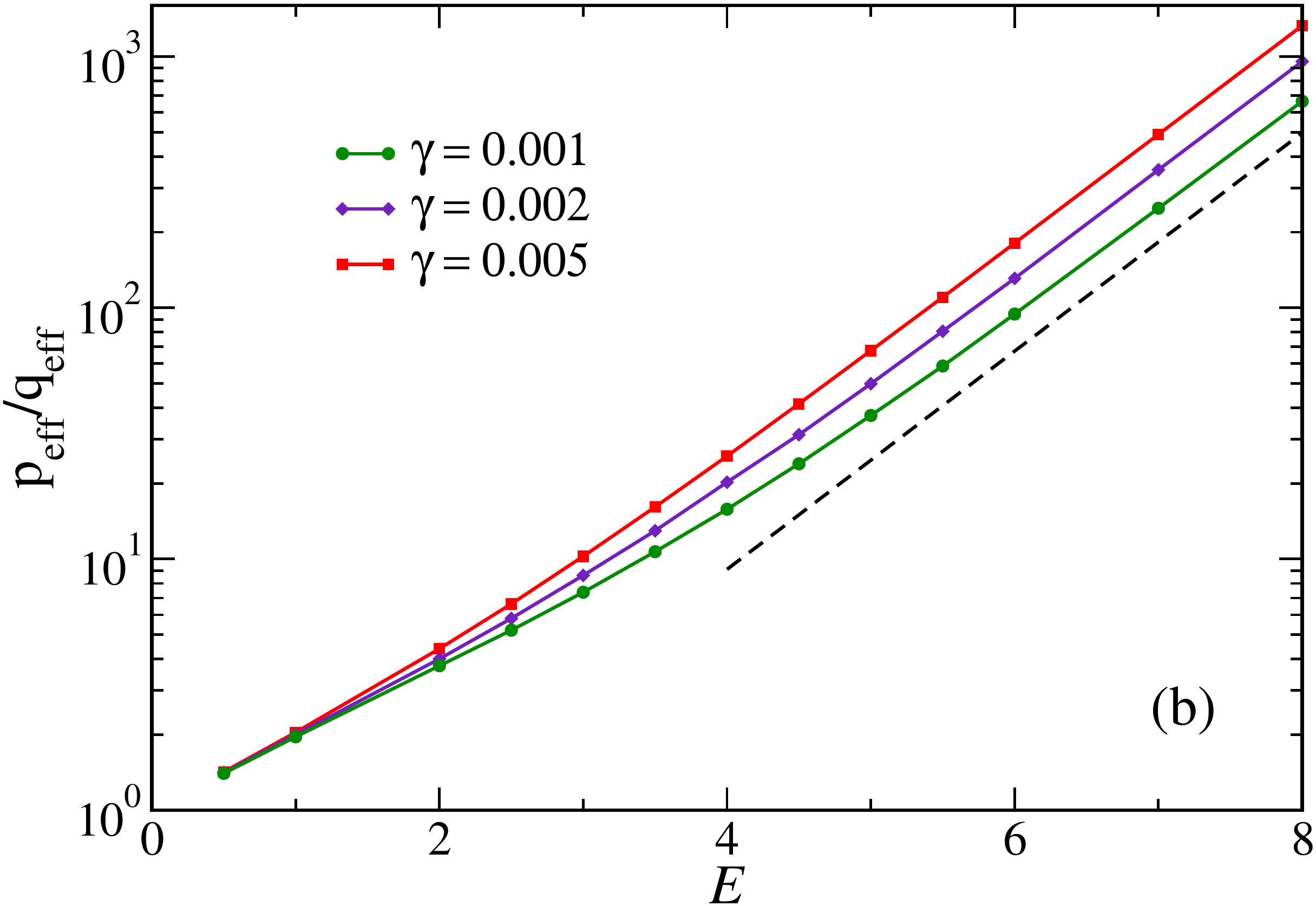} 
 \caption{ Variation of $(p+q)_{\text{eff}}$ (a) and  $p_{\text{eff}}/q_{\text{eff}}$ (b) with field $E$ for different values of $\gamma.$  Here $\rho=0.2.$ The dashed line in (b) corresponds to $e^{E}.$ }
 \label{fig:pqeff}
\end{figure}

\begin{figure}[t]
 \centering
 \includegraphics[width=7 cm]{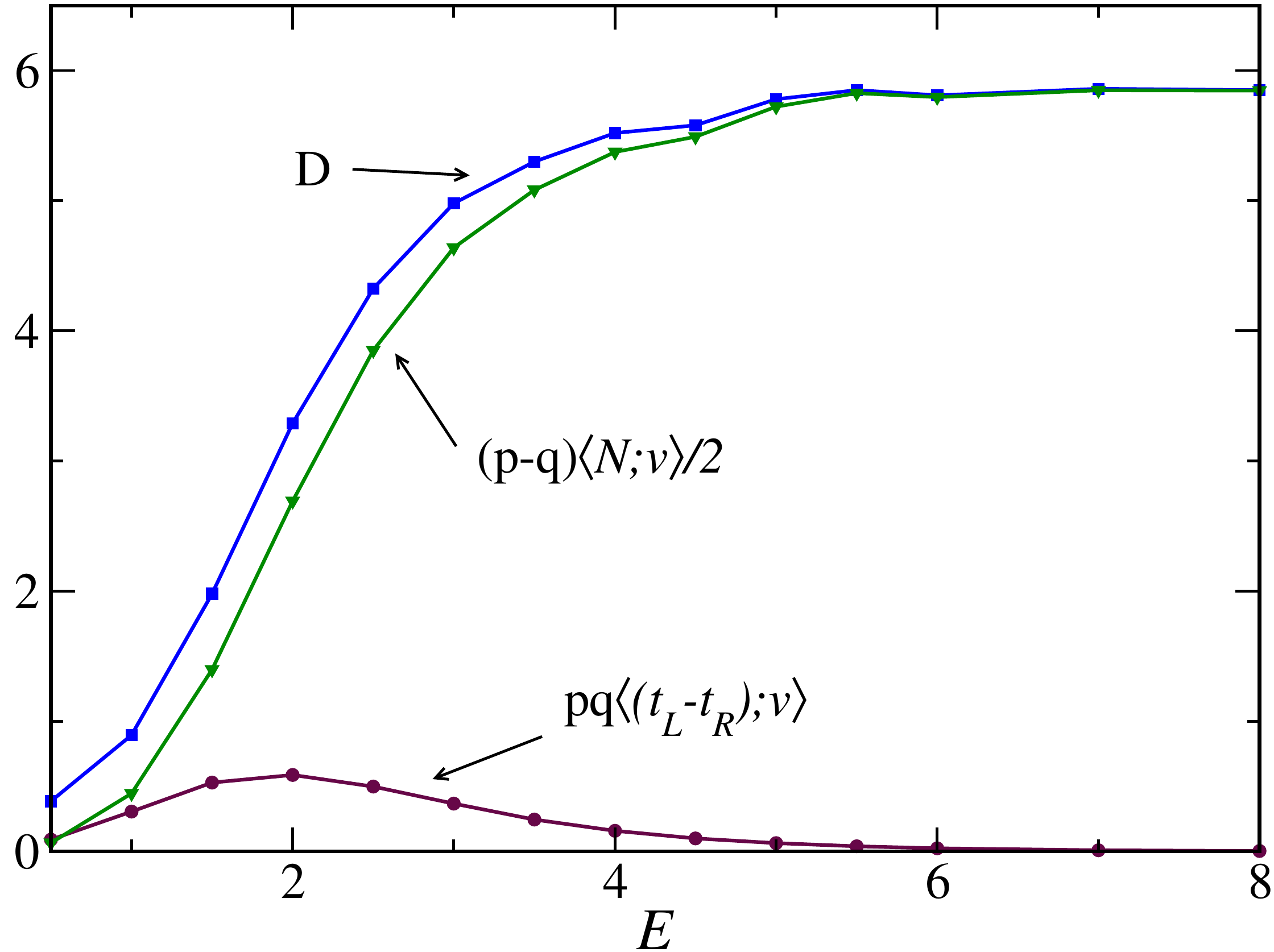}
 \caption{The different terms in the response formula (\ref{eq:djdE}) with the first and last terms coinciding for $E>4$. Here $\gamma= 0.001$ and $\rho=0.2.$   }
 \label{fig:terms}
 \end{figure}

  Note that the escape rate of the walker defined through (\ref{eq:star}) equals  $ p_{\text{eff}} + q_{\text{eff}} = \la N\ra/t $ as follows from comparing  (\ref{eq:vpqeff}) with (\ref{eq:Jpq}).  Following the writing of (\ref{eq:vr}), we then get the exact identity
 \bea
\langle v\rangle = p_{\text{eff}} - q_{\text{eff}} = \frac{p_{\text{eff}}/q_{\text{eff}} - 1}{p_{\text{eff}}/q_{\text{eff}} + 1}\,\la N\ra/t \label{eq:v_nav}
\eea
where
$$
\frac{p_{\text{eff}}}{q_{\text{eff}}} = e^{\beta_{{\text{eff}}}E}\,\quad \mbox{ for } \beta_{{\text{eff}}} = 1 + \frac 1{E}\log\frac{t-\langle t_R\rangle}{t-\langle t_L\rangle} 
$$
with Fig.~\ref{fig:pqeff}(b) showing indeed that $\beta_{{\text{eff}}}\simeq 1$ for large $E$.  It is therefore the decrease of $\langle N\rangle$ with large $E$ that decides the negative differential conductivity.\\
Continuing with (\ref{eq:ree}) the response of the average speed $\la v \ra$ is the differential mobility
\begin{equation}
\mu(E,\gamma) = \frac \id {\id E} \la v\ra = \underbrace{\frac 1{2t}\, \la J;J \ra}_D - pq\, \la (t_L-t_R); v\ra - \frac 1{2} (p-q)\, \la N;v \ra \label{eq:djdE}
\end{equation}
The first term is nothing but the diffusion constant $D=D(E,\gamma)$ for the charged particle which has its origin in the entropic contribution to the response. The other so called frenetic terms are truly nonequilibrium input and come from the time--symmetric part of the action; see \ref{ap}. We  emphasize that this response relation considers only the mobility of the charged particle to the change in field $E\rightarrow E +\id E$ and holds true for any value of $\gamma$ including zero. When $E=0$ and for no matter what $\gamma$ the mobility equals the diffusion constant (Sutherland--Einstein relation). The effects of $\gamma$ and $E$ are implicitly present in the expectations $\langle\cdot\rangle$ because the whole dynamical process depends on $\gamma$ and $E$. The different terms appearing in the response formula (\ref{eq:djdE}) are plotted against $E$ in Fig.\ref{fig:terms} for $\gamma=0.001,$ for which the negative response is most prominent. Note that the diffusion and the last term in (\ref{eq:djdE}
) approach each other for large field so that the response becomes almost equal to  $-pq\,\la (t_L-t_R); v\ra$:
   $$
   \mu(E,\gamma) \simeq (e^{E/2} + e^{-E/2})^{-2}\,\la (t_R-t_L); v\ra$$
 In the light of that formula, negativity of $\mu(E,\gamma)$ around $\gamma=0$ is then due to the strong positivity of the correlation 
$\la (t_L-t_R); J\ra$ between the difference $t_L-t_R$ in dwelling times near obstacles (left {\it versus} right) and the current $J$.
The picture that emerges is that for large $E$ and small $\gamma$ the charged particle traces a path keeping to the immediate right of the obstacles while for large $E$ and large $\gamma$ the charge waits some time left of the obstacles.  It thus resembles the strategy of a pedestrian who hurries to cross a square with randomly moving walkers; when these move fast, the pedestrian can afford to keep going straight and to wait a bit in front of each walker when blocked --- when the walkers move slow, the pedestrian will prefer to slalom around them.

It is also possible to express the current in nonlinear response around $E=0$.  The same type of computations are involved as in the above; see again \ref{ap}. 
We then get $\langle J\rangle = \alpha E -\delta E^3 + O(E^5)$ with 
\bea
\hspace*{-1.8 cm}\alpha = \frac 1{2}\langle J;J\rangle_0>0,\quad \delta = \frac 1{48} [3 \la N ; J^2 \ra_0 - \frac 34 \la (t_L-t_R)^2; J^2\ra_0 -(\la J^4 \ra_0 - 3 \la J^2\ra^2_0 )]\label{exp}
\eea
in terms of expectations $\langle\cdot\rangle_0$ at $E=0$ and where the last term $\la J^4 \ra_0 - 3 \la J^2\ra^2_0$ equals the fourth cumulant of $J$. Presence of a negative differential response would basically mean that $\delta$ is positive.
Measuring the positivity of $\delta$ from numerical simulations for  small $\gamma$ turns out however to be very difficult. That appears compatible with difficulties in computing (super-)Burnett coefficients, \cite{Burnett}. In general for characterizing a nonequilibrium regime, linear response around a stationary process possibly far from equilibrium is more useful than nonlinear response in expansion around equilibrium.

\subsection{Heuristics}

As mentioned in the introduction there are two regimes, a quasi--stationary regime where for small  $\gamma$ the obstacles move much slower than the charged particle and a similar behaviour to $\gamma=0$ is seen, and a fluid regime for large $\gamma$ where the charged particle moves as in an equilibrium fluid with background density $\rho$.\\

The fluid regime is characterized by $\gamma\gg 1$ where the effective parameters become  $ p_{\text{eff}}  = (1-\rho)p,  q_{\text{eff}}  = (1-\rho)q$ as indeed recovered numerically.\\ 
The quasi--static regime is specified by the relaxation time $\tau(E,\gamma)$ of the charged particle being much smaller with respect to $\gamma^{-1}$.  There a similar behavior as for $\gamma=0$ is observed, including the negative differential mobility $\mu(E,\gamma)$. If, with $E>E_c(0), \tau(E,\gamma)\ll \gamma^{-1}$, then $\mu(E,\gamma)<0$.  Being optimistic we could put at first $\tau(E,\gamma)=\tau(E_c(0),0)$ for small $\gamma$ to estimate the critical $\gamma_c$ from equating
$$
\tau(E_c(0),0) = \frac 1{\gamma_c}
$$
Since the numerics show $E_c(0) \simeq 1.25$ and $\tau(E=1.25,\gamma=0) \sim 10^4$, we would get $\gamma_c \simeq 10^{-4}$ while the simulations give $\gamma_c \simeq 10^{-3}$ which is off by one order of magnitude.  The reason is probably that for the relevant values of $E>E_c(0)$ and $\gamma$ the relaxation time $\tau(E,\gamma)$ is still one order of magnitude smaller than $\tau(E_c(0),0)$. \\ 

There is thus a critical $\gamma_c$ with for $\gamma<\gamma_c$ the existence of a finite $E_c(\gamma)$ with negative differential mobility $\mu(E,\gamma)$ for a range of $E>E_c(\gamma)$, while $\mu(E,\gamma) >0$ for all $E$ if $\gamma > \gamma_c$. In what follows we provide exact results for some analogous toy models to further support the above numerical evidence and this scenario in particular.

\section{Simplified random walker models}\label{sec:one}

An exact solution of the Lorentz model in a dynamic environment is of course very difficult to obtain when the time-scales of particle {\it vs.} environment do not diverge, $\gamma^{-1}\neq 0$. Yet the very similar phenomenon of mobility transition  can be exactly obtained for a number of much simpler quasi--one--dimensional random walkers.\\

\begin{figure}[t]
 \centering
 \includegraphics[width=12 cm,bb=0 0 528 128]{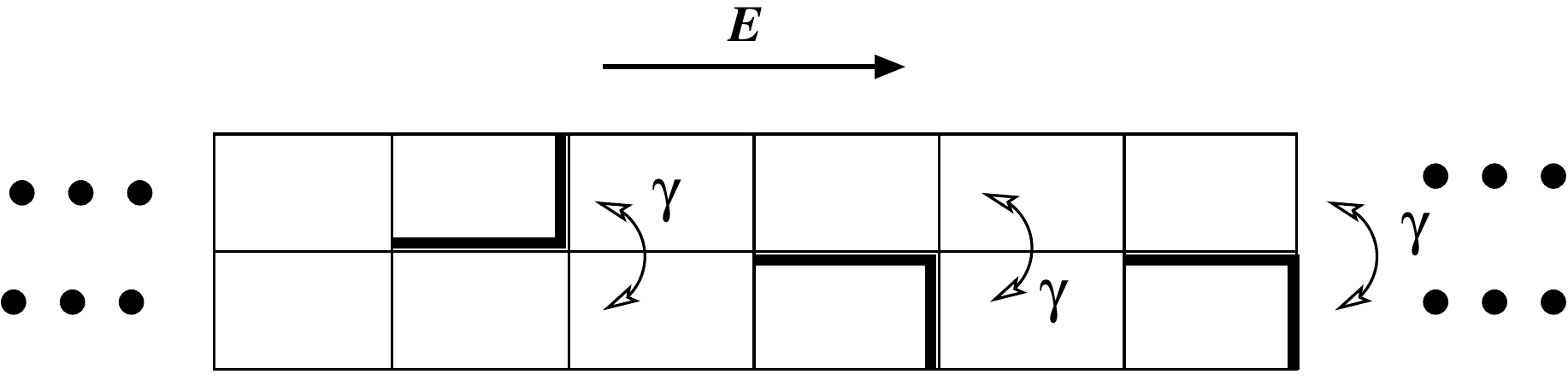}
 \caption{Schematic diagram of the two lane model of negative mobility}
 \label{fig:2lane_cartoon}
\end{figure}

To start we can consider a biased random walker on two lanes, see Fig.~\ref{fig:2lane_cartoon}, quite similar to the model in \cite{zia}. Motion in the horizontal direction is biased as in (\ref{eq:cho}) for external field $E$ and the vertical transitions (between the lanes) happen at rate $\frac 12$ if there is no wall (thick horizontal line in Fig.~\ref{fig:2lane_cartoon}).  The vertical wall can either stand in the upper or in the lower lane blocking (closed) or unblocking (open) the horizontal transitions. To keep the analogue with the Lorentz model of above, we imagine a rate $\gamma$ of transition for that wall (obstacle) moving up and down. 

\begin{figure}[t]
 \centering
 \includegraphics[width=8 cm]{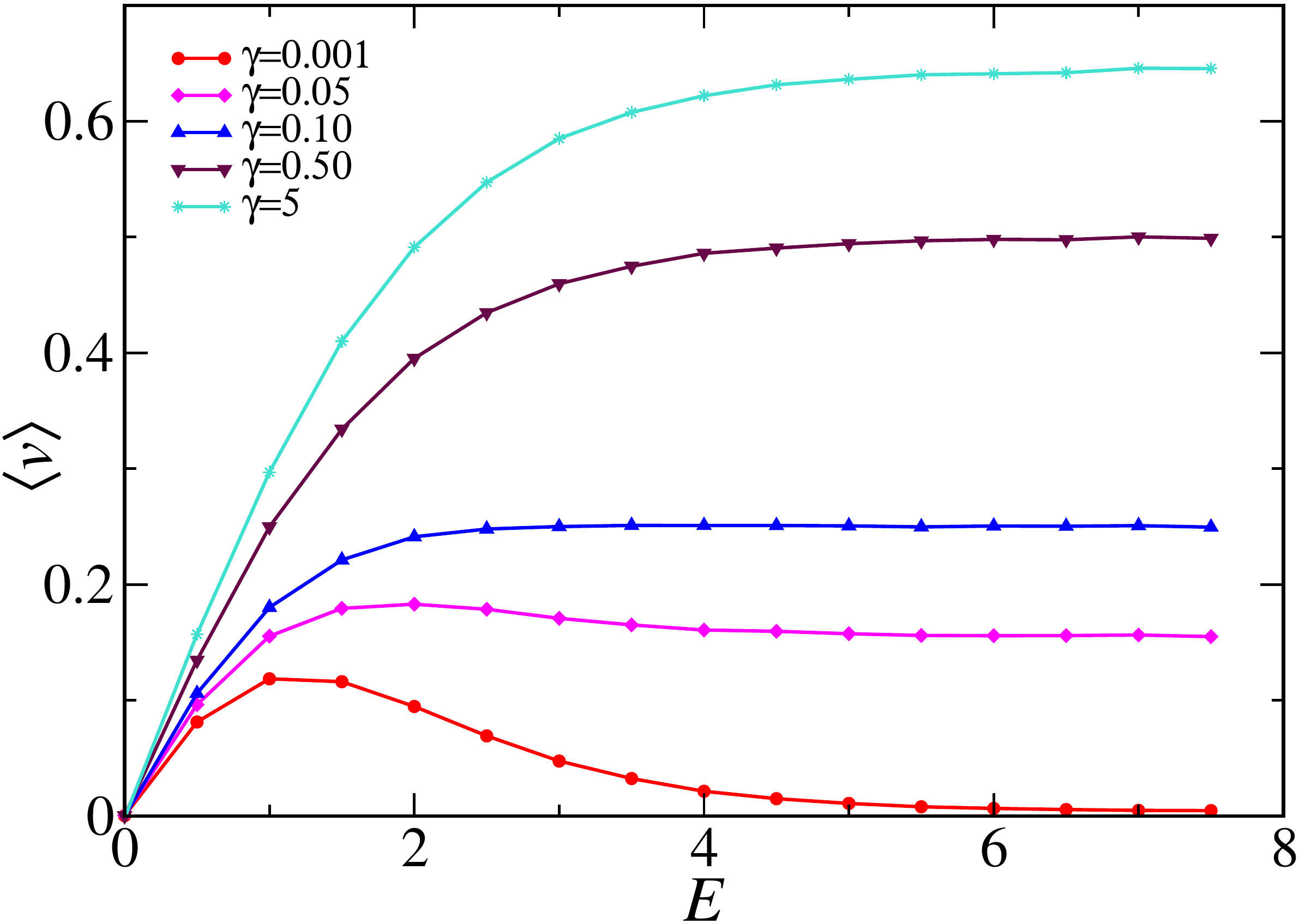}
 \caption{Average speed $\la v \ra$ for different values of $\gamma$ for the 2-lane model.}
 \label{fig:2lane_cur}
\end{figure}

For $\gamma=0$ we have a static environment and an exact solution can be easily obtained for any specific choice of obstacle configuration. For example, when all the vertically oriented obstacles are in the lower lane then the average speed is (see also \cite{zia,neg_resp})
\bea
\la v \ra^{\gamma=0} = \frac {2 \tanh (E/2)}{(3+e^{E})}
\eea
with a regime for the field $E$ with negative differential mobility. More precisely and in the notation of the previous sections, $E_c(0) \simeq 1.34.$\\
At the other extreme, for $\gamma\uparrow \infty$ (fluid limit) we obtain a monotonically increasing current $\langle v\rangle^{\gamma\uparrow \infty} = \frac 23 \tanh(E/2).$  Here, as in the Lorentz model with moving obstacles, there is a transition for a region of external fields $E$ where the differential mobility $\mu(E,\gamma)$ starts from being negative to become positive as  $\gamma$ increases;  see Fig. \ref{fig:2lane_cur}. The threshold  which separates the two regimes is estimated to be $\gamma_c \simeq 0.1.$ \\

\begin{figure}[b]
 \centering
 \includegraphics[width=12 cm]{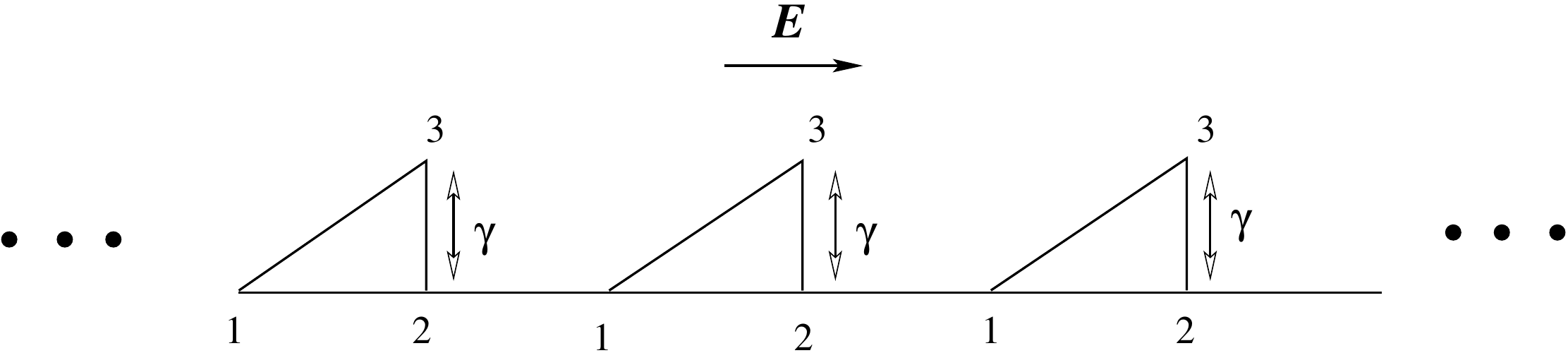}
 \caption{ Schematic representation of the 1D toy model for negative mobility}
 \label{fig:3st_model}
 \end{figure}

This model can still be simplified further without any essential changes.   We skip some steps in the simplification to arrive at a truly one--dimensional lattice as in Fig.~ \ref{fig:3st_model}. \\ 
All jumps towards the right occur with rate $p.$ This includes jumps $1 \to 2,$ $1\to 3$ and $2 \to 1$ in the forward direction. The  rate for the   jump in the corresponding backward directions is $q.$ The vertical jump rate, {\it i.e.}, between states 2 and 3 is symmetric and is denoted by $\gamma.$ 
 
In the stationary state, the average current and differential mobility can be calculated explicitly. Following the choice (\ref{eq:cho}) where $p+q=1,$ the steady state probabilities of being in states $1,2$ and $3$ are
$$
\rho_1 = \frac1Z [q+\gamma(1+q)], \quad \rho_2 = \frac1Z [q +\gamma(1+p)], \quad \rho_3 = \frac1Z [p +\gamma(1+p)]
$$
with  $Z=p(\gamma -1)+4\gamma+2.$ The average current per unit time is then given by,
\bea
\la v \ra &=& p \rho_2 - q \rho_1 \cr
&=&\frac{(p-q)(q+2\gamma)}{p(\gamma -1)+4\gamma+2} \label{eq:v_toy}
\eea
For $\gamma=0$  we find $\la v \ra^{\gamma=0} = \frac{q(p-q)}{p+2q}$ which vanishes for large $E.$
For non--zero $\gamma$ the average speed saturates to $\frac{2\gamma}{5\gamma+1}$ for $E \uparrow \infty.$

\begin{figure}[t]
 \centering
 \includegraphics[width=7.5 cm]{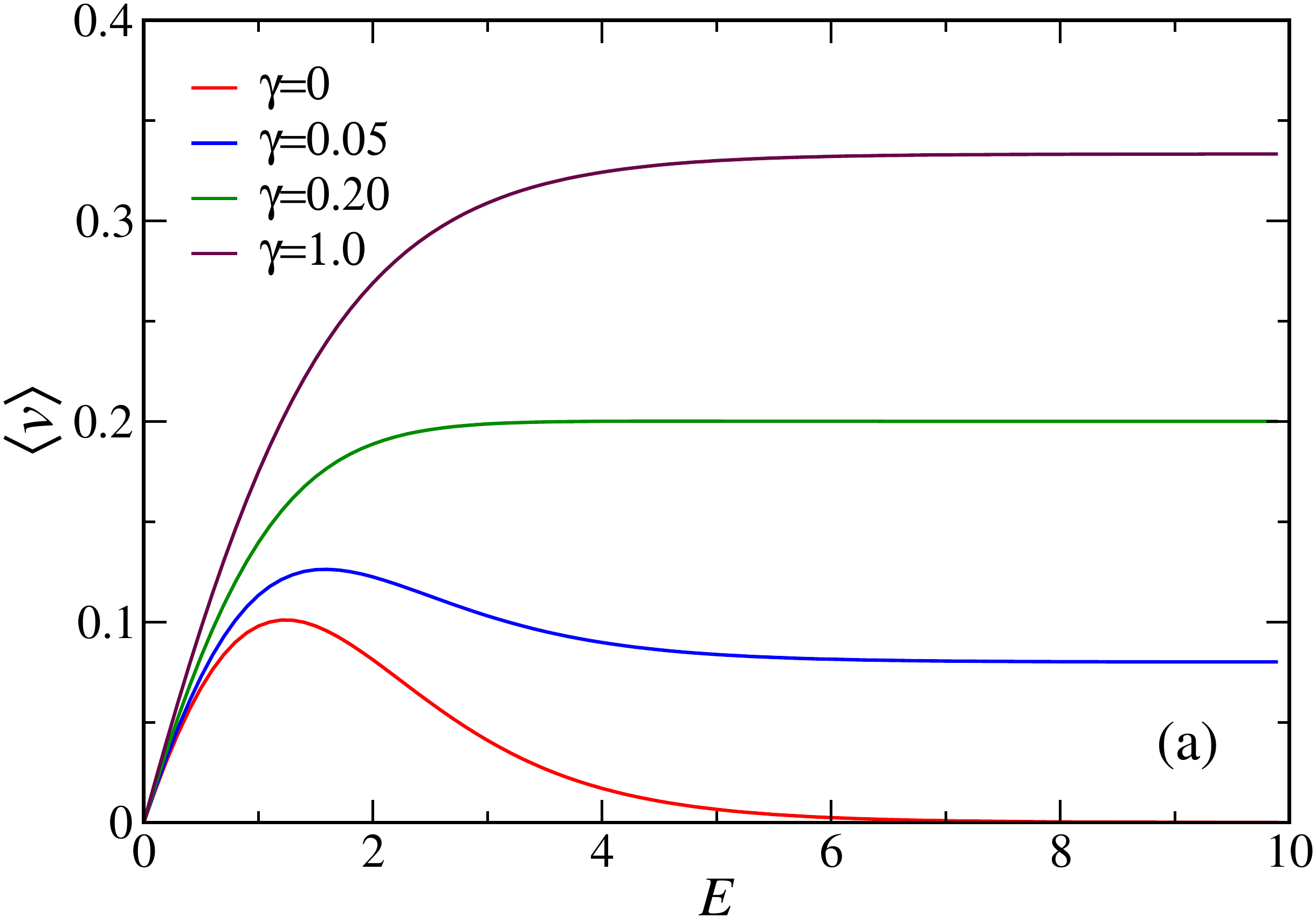}\hspace*{0.2 cm} \includegraphics[width=7.2 cm]{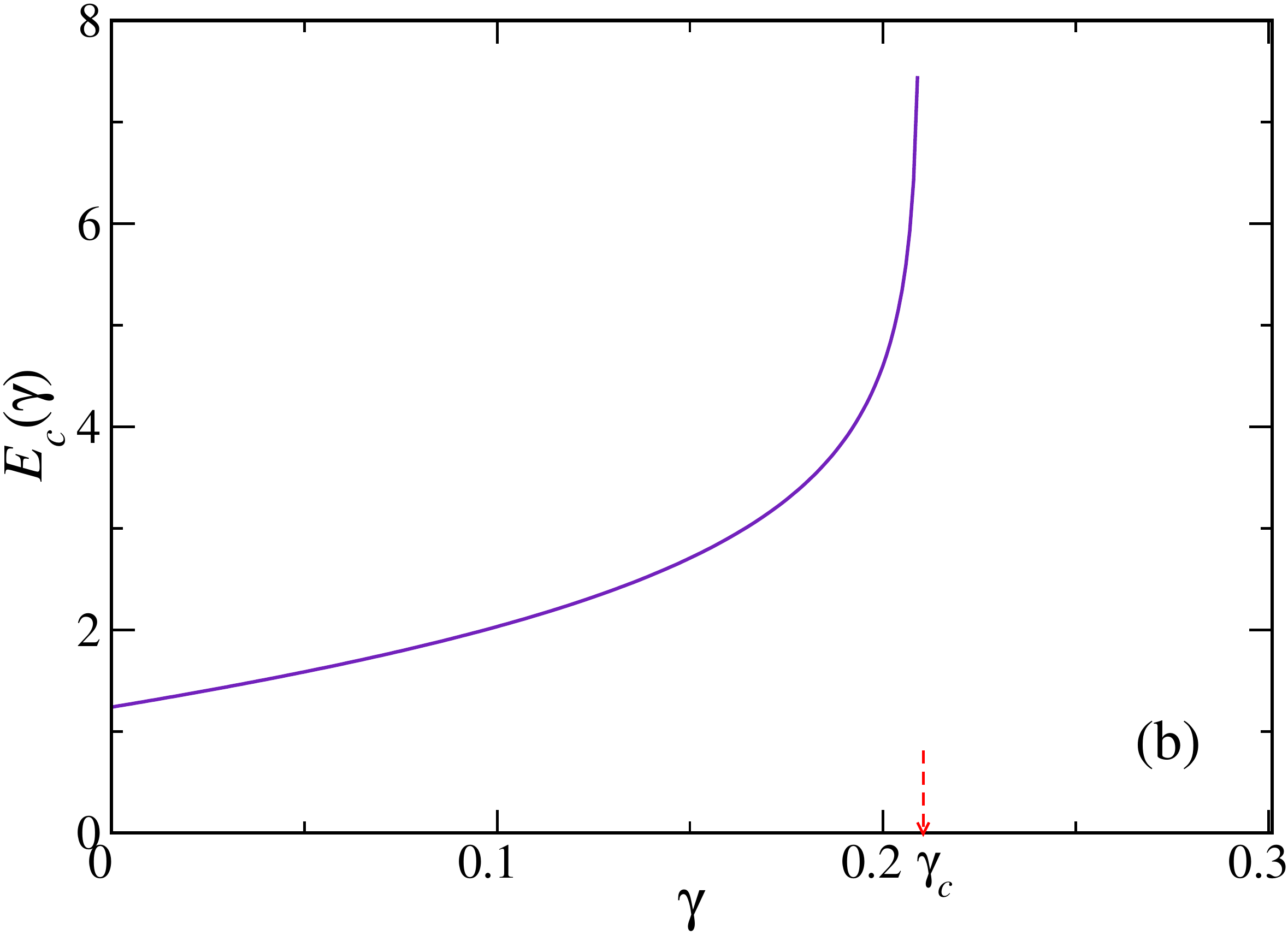}
 \caption{Toy model : (a) The current {\it versus} field $E$ from (\ref{eq:v_toy}). (b) $E_c(\gamma)$ {\it versus} $\gamma;$ giving no real solution beyond $\gamma_c \simeq 0.21.$}
 \label{fig:3st_cur}
\end{figure}

The plot in Fig.~\ref{fig:3st_cur}(a) shows exactly the same qualitative behaviour as for the driven Lorentz model in the previous section, compare with Fig.~\ref{fig:r0.2}(a).  It shows a transition in the sign of the differential mobility from negative for small $\gamma$ (quasi-static regime) and $E$ sufficiently large, to being always positive in the fluid limit (large $\gamma$).  Note also that $E_c(0)=\log(1+ \sqrt{6})\simeq 1.24$ and $\gamma_c = \frac{\sqrt{73}-1}{36} \simeq 0.21$ can be computed from putting $
\frac{\id\la v\ra}{\id E} =0$:
the solution is
\bea
E_c(\gamma)= \log \frac{1 + 11 \gamma + 18 \gamma^2 + \sqrt{6(1 + 6 \gamma + 11 \gamma^2 + 6 \gamma^3)}}{1 - \gamma - 18 \gamma^2}
\eea

Fig.~\ref{fig:3st_cur}(b) shows a plot of $E_c(\gamma)$ rapidly increasing with $\gamma$ and diverging at the critical $\gamma_c$.

\section{Summary}

We have studied linear response of a single particle under nonequilibrium driving in contact with a randomly evolving environment, parametrized by rate $\gamma$. We have shown that in the quasi--stationary regime, $i.e.$ when the environment is slow enough, the differential mobility of the particle can become negative for large driving. On the other hand, for fast relaxing environment $\gamma \uparrow \infty$ the system reaches a fluid limit with positive differential mobility for all values of the driving field. Thus one can expect a transition at a finite $\gamma_c$ from `negative mobility' to `positive mobility.'
Numerical evidence of this transition is provided with the example of the 2-dimensional Lorentz gas with moving obstacles. We have also illustrated this transition with a number of exactly solvable toy models.

\appendix
\section{From path--integration to response}\label{ap}
For various computations of response it is useful to apply a path-formalism \cite{fdr,fren}.  The starting point is the dynamical ensemble giving the probability distribution over all possible paths $\omega$ in configuration space over times $[0,t]$, formally
$$
P_E[\omega] = e^{-{\cal A}(\omega)} \,P_0[\omega]
$$
where we compare the path-probabilities in an external field $E$ (left-hand side) with respect to the ensemble at zero-external field (in right-hand side).
The action ${\cal A}(\omega)$, function of the path $\omega$ over time interval $[0,t]$, is for the driven lattice Lorentz model of Section \ref{sec:log} given by
\bea
- {\cal A}(\omega) &=& N_\rightarrow \log p + N_\leftarrow \log q + p(t-t_R) + q(t-t_L)\label{eq:lod}
\eea
where we have ignored field $E$-independent terms ($p,q$ depend on $E$). The response is governed by the derivative $\id {\cal A}/\id E$, since
\begin{eqnarray}
\langle O \rangle_E  &=& \langle O(\omega)\,e^{-{\cal A}(\omega)}\rangle_0 \cr
\frac{\id}{\id E}\langle O \rangle_E &=& -\langle O(\omega)\,\frac{\id {\cal A}}{\id E}(\omega)\rangle_E\label{eq:deri}
\end{eqnarray}
for a general path-observable $O$ on $[0,t]$ (not depending on $E$).
Doing as (\ref{eq:deri}) we get from (\ref{eq:lod}) that
$$
\frac {\id}{\id E} \la O \ra_E = \frac\beta 2 \la J;O \ra_E   - p^\prime \la (t-t_R);O \ra_E - q^\prime \la (t-t_L);O \ra_E + \frac 12 \left(\frac {p^\prime}p + \frac {q^\prime}q \right) \la N ; O \ra_E
$$
from which (\ref{eq:ree}) follows when plugging in (\ref{eq:cho}) (and leaving out the subscript $E$).\\
For further interpretations it is useful to make the decomposition ${\cal A} = {\cal D}-S/2$ where ${\cal D}$ is (twice) the time--symmetric component and $S$ is antisymmetric under time--reversal. These are called the  entropic ($S$) and the frenetic (${\cal D}$) components of the action and for Section \ref{sec:log} they equal, respectively,
\bea
S &=& (N_\rightarrow -N_\leftarrow) E = JE \cr
{\cal D} &=& p(t-t_R) + q(t-t_L) - \frac 12 (N_\rightarrow + N_\leftarrow) \log pq \cr
 &=& p(t-t_R) + q(t-t_L) + N\, \log (e^{E/2}+e^{- E/2}) \label{eq:DS}
\eea
These are again path--observables but they depend on the field $E$; we recognize $S$ as the path--dependent entropy production (linear in $E$) and ${\cal D} $ as a measure of dynamical activity (nonlinear in $E$); recall that $N$ is the \underline{total} number of jumps along the $x-$axis while $J$ is the \underline{net} number in the positive $x-$direction.\\

For the nonlinear response around detailed balance ($E=0$) we proceed similarly. Taking $O=J$ to be the current in (\ref{eq:deri}) we derive
\begin{eqnarray}
\frac{d \la J \ra}{dE} &=&   -\la {\cal A}' J\ra, \;
\frac{d^2 \la J \ra}{dE^2} = \la (-{\cal A}''+ ({\cal A}')^2) J \ra, \nonumber\\ 
\frac{d^3 \la J \ra}{dE^3} &=& \la (-{\cal A}'''+3{\cal A}'{\cal A}''-({\cal A}')^3)  J \ra \label{eq:Aprime}
\end{eqnarray}
where the primes denote $E-$derivatives.
Inserting here ${\cal A} = {\cal D} -\frac 12 S$ with (\ref{eq:DS}), we get at $E=0$
\bea
\left.\frac{\id \la J \ra}{\id E}\right|_0= \frac 12 \la JS' \ra_0 - \la J {\cal D}' \ra_0 = \frac 12 \la J^2 \ra_0 = \frac 12 \la J;J \ra_0
\eea
The second term cancels as ${\cal D}'$ is symmetric under time--reversal whereas $S'=J$ is antisymmetric. Similarly,
\bea
\left.\frac{\id^2 \la J \ra}{\id E^2}\right|_0 = - \la {\cal D}''J \ra_0 + \la ({\cal D}')^2J \ra_0 + \frac 14 \la (S')^2 J \ra_0 - \la {\cal D}' S' J \ra_0 \label{eq:d2J0}
\eea
Each of the terms vanishes individually from symmetry considerations --- the first three from time reversal symmetry and the last one from reflection (left-right) symmetry.\\
The third derivative in (\ref{eq:Aprime}) is also calculated using (\ref{eq:DS})
\bea
\left.\frac{\id^3 \la J \ra}{\id E^3}\right|_0 &=& -\la {\cal D}''' J\ra_0  + 3 \la {\cal D}' {\cal D}'' J \ra_0    - \frac 32 \la {\cal D}'' S' J\ra_0 - \la ({\cal D}' - \frac 12 S')^3 J \ra_0 \cr
&=& \frac 18 \la J^4 \ra_0 + \frac 38 \left[\frac 14 \la (t_L-t_R)^2 J^2 \ra_0 - \la N J^2 \ra_0 \right] \n
\eea
which can be expressed as sums of cumulants and covariances: 
\bea
\left.\frac{\id^3 \la J \ra}{\id E^3}\right|_0 = \frac 18 [ \la J^4 \ra_0 - 3 (\la J^2\ra_0)^2 + \frac 34 \la (t_L-t_R)^2 ; J^2 \ra_0 - 3 \la N ;J^2 \ra_0] \label{eq:cov}
\eea
That yields the expansion (\ref{exp}) to third order in $E$.\\

\end{document}